\documentclass[aps,prapplied,10pt,twocolumn,superscriptaddress,amsmath,amssymb]{revtex4-2}

\usepackage{graphicx}
\usepackage{multirow}
\usepackage[usenames]{color}

\usepackage{hyperref}

\usepackage{placeins}
\usepackage{soul}
\begin{document}

\title{Expanding the operational temperature window of a superconducting spin valve}

\author{A.~A.~Kamashev}
\affiliation{Zavoisky Physical-Technical Institute, FRC Kazan
Scientific Center of RAS, 420029 Kazan, Russia}

\author{N.~N.~Garif'yanov}
\affiliation{Zavoisky Physical-Technical Institute, FRC Kazan
Scientific Center of RAS, 420029 Kazan, Russia}

\author{A.~A.~Validov}
\affiliation{Zavoisky Physical-Technical Institute, FRC Kazan
Scientific Center of RAS, 420029 Kazan, Russia}

\author{V.~Kataev}
\affiliation{Leibniz Institute for Solid State and Materials
	Research, Helmholtzstr. 20, D-01069 Dresden, Germany}

\author{A.~S.~Osin}
\affiliation{L.~D.\ Landau Institute for Theoretical Physics RAS, 142432 Chernogolovka, Russia}

\author{Ya.~V.~Fominov}
\affiliation{L.~D.\ Landau Institute for Theoretical Physics RAS, 142432 Chernogolovka, Russia}
\affiliation{Laboratory for Condensed Matter Physics, HSE University, 101000 Moscow, Russia}

\author{I.~A.~Garifullin}
\affiliation{Zavoisky Physical-Technical Institute, FRC Kazan Scientific Center of RAS, 420029 Kazan, Russia}

\date{\today}

\begin{abstract}
To increase the efficiency of the superconducting spin valve (SSV), special attention should be paid to the choice of ferromagnetic materials for the F1/F2/S SSV multilayer. Here, we report the preparation and the superconducting properties of the SSV heterostructures where Pb is used as the superconducting S layer. In the magnetic part of the structure, we use the same starting material, the  Heusler alloy Co$_2$Cr$_{1-x}$Fe$_x$Al$_{y}$, for both F1 and F2 layers. We utilize the tunability of the magnetic properties of this alloy, which, depending on the deposition conditions, forms either an almost fully spin-polarized half-metallic F1 layer or a weakly ferromagnetic F2 layer. We demonstrate that the combination of the distinct properties of these two layers boosts the generation of the long-range triplet component of the superconducting condensate in the fabricated SSV structures and yields superior values of the triplet spin-valve effect of more than 1\,K and of the operational temperature window of the SSV up to 0.6\,K.
\end{abstract}

\maketitle

\section{Introduction}

The constantly growing  interest in superconducting spintronics (SS) \cite{Linder} is due to the fact that its logic elements are based on the control of the spin of an electron rather than of its charge. In this regard, SS devices can operate at a higher speed and with a greater energy efficiency as compared to semiconductor electronics devices, where only the electron charge is controlled (see, e.g., Refs. \cite{Zutic,Efetov_Springer_2007,Efetov_Springer_2013}).
Over the past three decades, various logic elements for SS have been proposed and investigated
(see, e.g., Refs. \cite{Ioffe,Feigelman,Buzdin1,Bergeret,Blamire,Linder,Eschrig,Ryazanov1999,Ryazanov2000,Ryazanov2001,Ryazanov2001a,Kontos}).
As a rule, the operation of modern SS devices makes use of the superconductor/ferromagnet (S/F) proximity effect resulting from the mutual influence of ferromagnetism on superconductivity and {\it vice versa} in an S/F thin layer heterostructure \cite{Buzdin1,Garifullin_obzor_2002,Demler1997,Linder2009,Gaifullin,Bergeret,Blamire,Linder,Eschrig,Stoutimore,Pugach2009}.

Two theoretical models of a superconducting spin valve (SSV) based on the S/F proximity effect were proposed at the end of the 1990s by Oh
{\it et al.} \cite{Oh} (the F1/F2/S model) and by Tagirov \cite{Tagirov} and Buzdin {\it et al.} \cite{Buzdin2} (the F1/S/F2 model). Both theories treat a trilayer SSV composed of one superconducting and two ferromagnetic layers where the superconducting critical temperature $T_c$ is controlled by the mutual orientation of the magnetizations of the F1 and F2 layers. $T_c$ of such SSV structure is expected to be larger for the antiparallel (AP) mutual orientation
than for the parallel (P) one. This is because the average exchange field of the two F layers acting on the Cooper pairs from the S layer is smaller for the AP configuration as compared to the P case. Actually, this conventional, so-called standard, SSV effect is only one of the possibilities. Under certain conditions, $T_c^\mathrm{AP}$ in an SSV can be smaller than $T_c^\mathrm{P}$,
which corresponds to the inverse SSV effect \cite{Fominov,Mironov2014}. If the magnitude of the SSV effect $\Delta T_c=\left| T_c^\mathrm{AP}-T_c^\mathrm{P}\right|$ exceeds the superconducting
transition width $\delta T_c$, it becomes possible to completely switch on/off the superconducting current in the SSV structure. The
SSV can then be used as a passive logic element in SS, for example, as a superconducting key or switcher. The main qualitative
criterion for evaluating the efficiency of the SSV is the width of the operational temperature window
$\Delta T_c^\mathrm{full}$ within which it is possible to switch between
the normal and superconducting states. Larger value of $\Delta T_c^\mathrm{full}$ increases the efficiency of the SSV.

Historically, the F1/S/F2-type SSV structures were the first ones that were fabricated and studied experimentally \cite{Gu,Moraru,Potenza}. However, the full SSV effect was not achieved there.
The first experimental realization of the full switching on/off of a superconducting current was attained on the F1/F2/S-type SSV structure where the full SSV effect was demonstrated in the  Fe1/Cu/Fe2/In multilayer \cite{Leksin2010}. For this structure, $\Delta T_c = 19\mathrm{\,mK} > \delta T_c \sim 7$\,mK was found yielding the operational temperature window of this SSV  $\Delta T_c^\mathrm{full}$ of the order 10\,mK. Following this first success different SSV structures with a variety of ferromagnetic and superconducting materials were investigated in the past years \cite{Blamire,Eschrig,Grein,Flokstra,Montiel,Banerjee,Gu2015,Pugach2022,Aarts2015,Leksin2011,Leksin2012,Leksin2015,Garifullin,Leksin2016,JMMM,Kamashev2019,Kamashev20191}.

The clue to increase the operational temperature window was found in the generation of the long-range triplet component (LRTC) of the superconducting condensate of the S layer in the F part that enabled an increase of $\Delta T_c^\mathrm{full}$ up to $0.3$\,K in Ref.\ \cite{Kamashev20191} and even up to $\sim 0.45$\,K in Ref.\ \cite{Aarts2015}. In this case, the SSV effect is maximized by switching between the parallel and orthogonal configuration of the magnetizations.

The occurrence of the LRTC was predicted by Bergeret {\it et al.}~\cite{Bergeret2} by analyzing the processes taking place during the penetration of the Cooper pairs from an S layer into an F part of the proximity structure  (see Refs.\ \cite{Bergeret,Linder,Eschrig} for review). According to the theory by Fominov {\it et al.} \cite{Fominov}, the presence of a $T_c$ minimum near orthogonal orientation of the F layers' magnetizations evidences the generation of the LRTC in the F1/F2/S-type SSV structures. Due to the generation of the LRTC, an additional channel arises for the leakage of Cooper pairs from the S to F part in the F1/F2/S structure. This can lead to a significant suppression of $T_c$ and, consequently, to an increase of the efficiency of the SSV structure. A number of recent studies are thus 
dealing with manifestations of the LRTC in SSV structures \cite{Leksin2011,Kamashev2019,Kamashev20191,Gu,Wu,Gu2015,Aarts2015,Halterman1,Halterman2,Halterman3,Halterman4,Halterman5}.

According to the results in Refs.\ \cite{JMMM,Aarts2015,Kamashev2019}, the magnitude of the operational temperature window $\Delta T_c^\mathrm{full}$ strongly depends on the specific properties of ferromagnetic materials used in an F1/F2/S structure.
Therefore, optimizing their choice can significantly increase $\Delta T_c^\mathrm{full}$. In this respect, the use of a weak ferromagnetic material
as the F2 layer appears to be promising. It may make possible to observe large values of $\Delta T_c$, since the F2 layer performs as a kind of a selective ``filter'' which exposes the S layer to the influence of the strong magnetization of the F1 layer {\em only} in the presence of the LRTC, by this making the S/F proximity effect highly tunable and pronounced.
Earlier we found that the Heusler alloy Co$_2$Cr$_{1-x}$Fe$_x$Al$_y$ (HA)
film deposited on a substrate kept at a temperature $T_\mathrm{sub} \sim 300$\,K  appears to be a  weak ferromagnet (HA$^\mathrm{RT}$) \cite{Kamashev2017}, and if used as an F2 layer it increases the $\Delta T_c$ value of the F1/F2/S SSV structure  compared to our previous results with a more conventional ferromagnetic material in the role of the F2 layer \cite{JMMM}.

In contrast, this HA film becomes practically half-metallic (HA$^\mathrm{hot}$) with the degree of the spin polarization (DSP) of its conduction band of $70-80$\,\% if prepared at $T_\mathrm{sub}\geq 600$\,K \cite{Kamashev2017}. The use of such a high DSP ferromagnetic material as the F1 layer may increase the magnitude of $\Delta T_c^\mathrm{full}$ since it allows a significant suppression of $T_c$ at the orthogonal orientation of the F1 and F2 magnetizations compared to the collinear one due to generation of the LRTC of the superconducting condensate.

In 2015, Singh {\it et al.} \cite{Aarts2015} discovered a giant magnitude of the triplet spin-valve
effect $\Delta T_c^\mathrm{trip}=T_c(\alpha=0^\circ)-T_c(\alpha=90^\circ) \geq 0.7$\,K, where $\alpha$ is the angle between the cooling field and the direction of the
applied magnetic field [$\alpha=0^\circ$ corresponds to the parallel and $\alpha=90^\circ$ corresponds to the perpendicular (PP) orientation of the F layers' magnetizations].
In that work the half-metal CrO$_2$ with 100\,\% DSP was used in the place of the F1 layer. We also obtained similar results for the F1/F2/S structures where HA$^\mathrm{hot}$ with a high DSP was used as the F1 layer \cite{Kamashev2019,Kamashev20191}.

In the present paper, we propose and realize a concept to exploit the ferromagnetic dualism of the HA material to maximize the magnitude of $\Delta T_c^\mathrm{full}$. 
For that we use within the same F1/F2/S structure this HA in a two-fold role, its HA$^\mathrm{hot}$ version as a half metal in the place of the F1 layer, and its HA$^\mathrm{RT}$ 
variant as a weak ferromagnet in the place of the F2 layer. With this type of SSV heterostructure we achieved the magnitude of the triplet spin-valve effect $\Delta T_c^\mathrm{trip}$ 
of more than 1 K and expand the operating temperature window $\Delta T_c^\mathrm{full}$ up to $\sim$ 0.6 K. Our results pave the way towards optimization of ferromagnetic materials in an F1/F2/S SSV structure together with the simultaneous simplification of its fabrication which should stimulate further experimental and theoretical studies of the interplay between ferromagnetism and superconductivity in F/S heterostructures.

\section{Samples}

The samples were grown within a closed vacuum
cycle using the deposition system from BESTEC GmbH composed of a load lock
station with vacuum shutters, a molecular beam epitaxy (MBE) chamber with a stationary vacuum of about $1\times10^{-10}$\,mbar and a magnetron sputtering chamber with a stationary vacuum of about $1\times10^{-9}$\,mbar. The prepared heterostructures have the following composition:
HA$^\mathrm{hot}$(20nm)/\allowbreak Al(4nm)/\allowbreak HA$^\mathrm{RT}$($d_{\mathrm{HA}^\mathrm{RT}}$)/\allowbreak Al(1.2nm)/\allowbreak Pb(60nm)/\allowbreak Si$_3$N$_4$ with the variable HA$^\mathrm{RT}$ layer thickness $d_{\mathrm{HA}^\mathrm{RT}}$ in the range from 1 to 5\,nm.
In this construction HA$^\mathrm{hot}$ and HA$^\mathrm{RT}$ play the roles of the ferromagnetic F1 and F2 layers, respectively; 
Pb(60nm) is an S layer; Si$_3$N$_4$ is a protective layer against oxidation. 

The Al layer sandwiched between the HA$^\mathrm{hot}$ (F1) and HA$^\mathrm{RT}$ (F2) layers is needed to decouple magnetizations of these two F layers. Its optimal thickness of 4 nm is chosen based on our previous 
experimental works in Refs. \cite{Leksin2010,Leksin2011,Leksin2012,Leksin2015,Garifullin}. The second Al layer deposited on top of the S layer is used to stabilize the superconducting properties of the SSV structures 
and, in particular, to prevent intermixing of the S layer with the adjacent F2 layer while not suppressing too much the superconducting $T_\mathrm{c}$. The thickness of this layer of 1.2 nm was chosen based on the detailed 
study of the influence of the thickness of this buffer layer on the properties of the SSV structures in Ref. \cite{Leksin2015}. The thicknesses of the Al layers are expected to be small compared to the corresponding coherence length, so that the Al layers only have a technological effect improving properties of the interfaces.
All materials used for evaporation had a purity of better than 4\textit{N} (99.99\,at.\,\%). The heterostructures were deposited on high quality single crystalline substrates MgO(001).
The Al and Pb layers were deposited in the MBE chamber using the e-beam technique. Other parts of the structure were prepared in the magnetron sputtering chamber using the DC sputtering technique for the deposition of the HA$^\mathrm{hot}$ and HA$^\mathrm{RT}$ layers and the AC sputtering technique for the deposition of the Si$_3$N$_4$ layer.
The deposition rates were as follows: 0.4\,{\AA}/s for HA$^\mathrm{hot}$ and HA$^\mathrm{RT}$,
0.5\,{\AA}/s for Al, 12\,{\AA}/s for Pb, and 1.8\,{\AA}/s for Si$_3$N$_4$ films.

\begin{figure}[t]
\includegraphics[width=0.7\columnwidth]{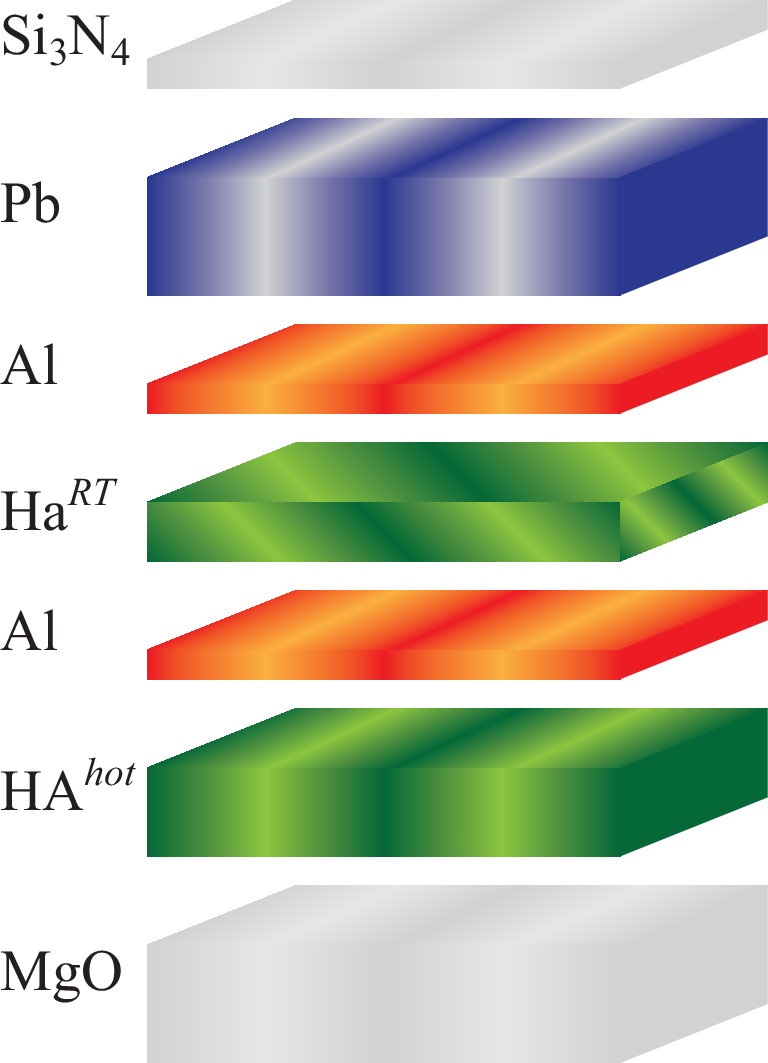}
\caption{Design of the prepared SSV heterostructures (see the text for details).}
\label{fig1}
\end{figure}

First, the substrates were fixed on a sample holder and transferred into the magnetron sputtering chamber through the load lock station for preparing HA$^\mathrm{hot}$ layers.
We used a specially designed
rotating wheel sample holder for preparing a set of samples with different layer thicknesses in a single vacuum cycle. When evaporating HA$^\mathrm{hot}$, the substrate temperature was kept at $T_\mathrm{sub} \sim 700$\,K to achieve the maximum  DSP of
the HA's conduction band since Husain {\it et al.} \cite{Husain} showed that the DSP
increases with increasing the $T_\mathrm{sub}$. After the preparation of the HA$^\mathrm{hot}$ layer,
the substrate was cooled down to room temperature and other layers of the heterostructure were deposited.
According to our previous study in Ref.\ \cite{Nano}, to improve the smoothness of the Pb layer the substrate temperature should be reduced down to $T_\mathrm{sub}\sim 150$\,K. Therefore, the top Al/Pb fragment was deposited at this temperature.
Finally, all samples were covered with a protective Si$_3$N$_4$ layer to prevent their oxidation. The design of the prepared heterostructures is presented in Fig.~\ref{fig1}.

At variance with our previously established design in Refs.\ \cite{Leksin2010,Leksin2011,JMMM,Leksin2016,Garifullin}, we did not use here CoO$_{x}$ as an antiferromagnetic bias layer that pins the  direction of the magnetization of the F1 layer necessary for the operation of the SSV. At a high substrate temperature of $\sim700$\,K required for the deposition of the F1 HA$^\mathrm{hot}$ layer, CoO$_{x}$ decomposes to ferromagnetic cobalt and loses its biasing property. Instead, we apply the field cooling procedure to bias the F2 magnetization as explained in the next Section.

\section{Experimental results}

The superconducting properties of the samples were measured using the standard four-point method with DC current. The current and voltage leads were attached to the sample by clamping contacts. A highly homogeneous vector electromagnet from Bruker Instruments was used for measurements in an external magnetic field. The sample was mounted in the $^4$He cryostat that was inserted between the poles of the electromagnet. This configuration allowed the rotation of the magnetic field in the plane of
the sample with an accurate control of the actual magnetic field acting on the sample. The magnetic field was measured with an accuracy of $\pm 0.3$\,Oe using a Hall probe. To avoid the occurrence of the out-of-plane component of the external field the sample plane position was always
adjusted relative to the direction of the external field with an accuracy better than $1.5^\circ$. The temperature of the sample was monitored by the Allen–Bradley resistor thermometer which is particularly sensitive in the temperature range of our interest. The critical temperature $T_c$ of the samples was defined as the midpoint of the resistivity transition curve.

The quality of the Pb layer can be estimated from the residual resistivity ratio $\mathrm{RRR}=R({\rm 300 K})/R({\rm 10 K})$.
For all samples RRR was in the range $15-20$, which indicates a high quality of the prepared Pb layer.

The thickness of the S layer $d_\mathrm{Pb}$ is a very important parameter for the efficient operation of the SSV structure. It should be sufficiently small to
make the whole S layer sensitive to the magnetic part of the system. Only in this case the mutual orientation of the magnetizations of the F1 and F2 layers
would affect the $T_c$. To determine the optimal $d_\mathrm{Pb}$, we used our previous results from Ref.\ \cite{Kamashev2017} where the $T_c(d_\mathrm{Pb})$ dependence was calibrated for the HA$^\mathrm{RT}$/Cu/Pb structures at a fixed thickness of the HA$^\mathrm{RT}$ layer.
The $T_c$ decreases rapidly upon reducing $d_\mathrm{Pb}$ down to 50\,nm and drops sharply below 1.4\,K at $d_\mathrm{Pb}\leq 25$\,nm.
Therefore, the optimal thickness range of the Pb layer lies in between 50 and 70\,nm. For this reason  the working thickness of the Pb layer for the samples studied in the present paper
was chosen as $d_\mathrm{Pb} = 60$\,nm.

The study of the magnetic properties of the samples was carried out using a vibrating sample magnetometer superconducting quantum interference device magnetometer. For that the magnetic hysteresis loops of the specially prepared single-layer samples of HA$^\mathrm{hot}$ ($d_{\mathrm{HA}^\mathrm{hot}}=20$\,nm) and HA$^\mathrm{RT}$ ($d_{\mathrm{HA}^\mathrm{RT}}=3$\,nm) with the same thicknesses as used in the SSV heterostructures were measured.
Otherwise, if HA$^\mathrm{hot}$ and HA$^\mathrm{RT}$ are combined in the F1/F2/S multilayer, it is hard to resolve their magnetic hysteresis loops since the signal comes from the layers the thicknesses of which differ by almost an order of magnitude.

\begin{figure}[h]
	\includegraphics[width=\columnwidth]{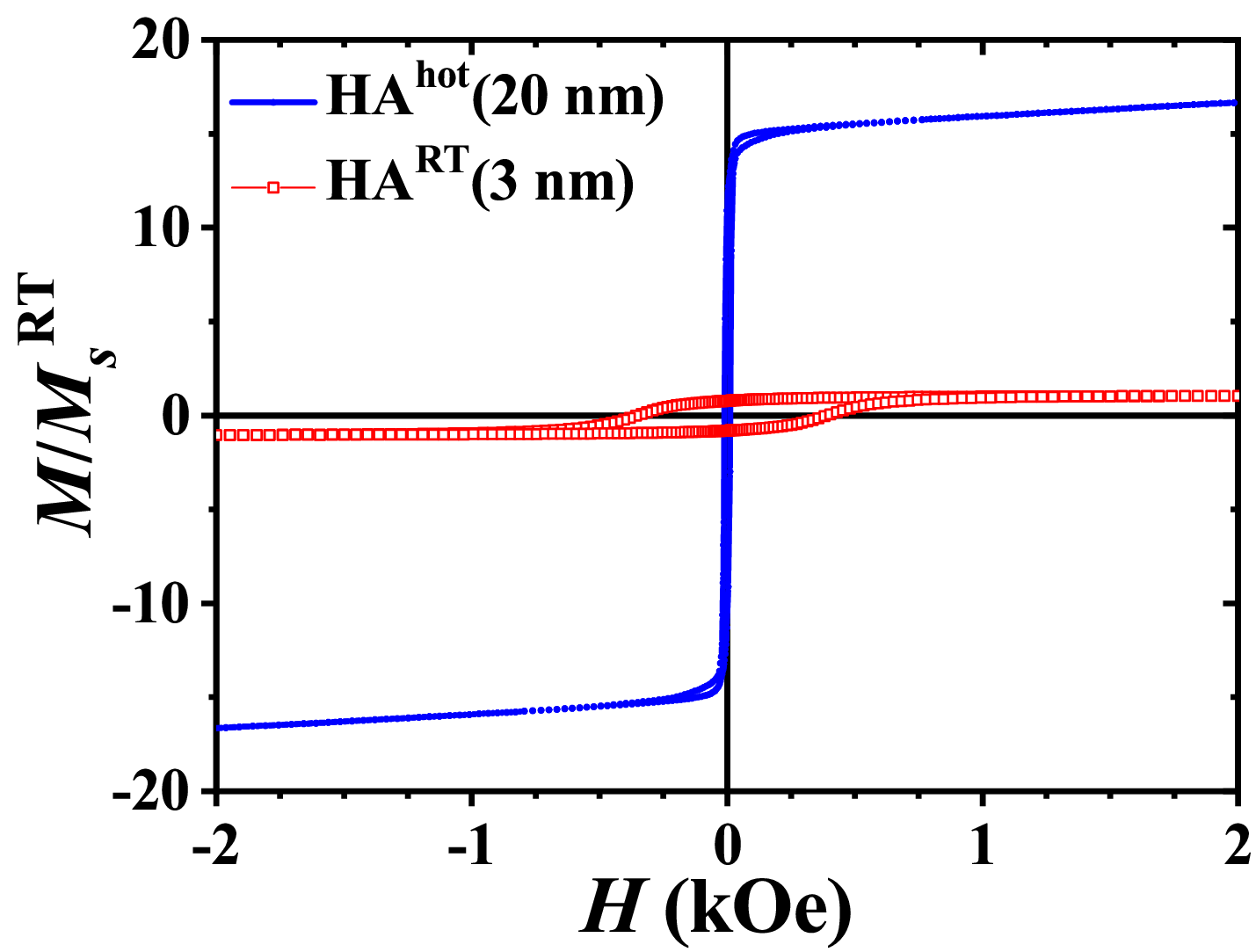}
	\caption{Magnetic hysteresis loops for the HA$^\mathrm{hot}$ ($d_{\mathrm{HA}^\mathrm{hot}}=20$\,nm) and HA$^\mathrm{RT}$ ($d_{\mathrm{HA}^\mathrm{RT}}=3$\,nm) single layer samples measured at $T = 30$\,K.	
	The magnetization curve of the HA$^\mathrm{RT}$ sample is scaled to unity whereas the respective curve of the HA$^\mathrm{hot}$ sample is scaled to the ratio of the absolute saturation magnetizations of the two layers $M_s^{\rm hot}/M_s^{\rm RT}$.} 
	
	\label{fig2}
\end{figure}

Obviously, there is also a significant difference in coercivities of the F1 layer (HA$^\mathrm{hot}$) and F2 layer (HA$^\mathrm{RT}$), see Fig.~\ref{fig2}. Magnetization of the F1 layer saturates at approximately 30\,Oe, which is comparable to its coercive field, whereas for the F2 layer the saturation of magnetization is reached at 1\,kOe and the coercive field is close to 0.5\,kOe.
It is worth noting that in still higher magnetic fields, the magnetization of the F1 layer continues to increase slightly possibly due to some magnetic inhomogeneity of the HA$^\mathrm{hot}$ layer. The difference in coercive forces can be first due to different preparation conditions
of the F1 and F2 layers. It can be assumed that the coercive field increases with decreasing $T_\mathrm{sub}$ because the density of any defects increases at lower deposition temperatures.
The second reason for the difference in the coercive forces can be different thicknesses of the  F1 and F2 layers. From general considerations, it can be assumed that with  decreasing the ferromagnetic film thickness the anisotropy field of the film increases,
which in turn leads to an increase of its coercive force (see, e.g., Ref. \cite{Leksin2009}).

The procedure for measuring $T_c$ of the SSV samples in an external magnetic field was as follows. Pinning of the magnetization of the F2 layer (HA$^\mathrm{RT}$) in a certain direction was achieved by cooling the sample in a magnetic field of about 8\,kOe  from room temperature down to the low operational temperatures.
The magnetization of the F1 layer (HA$^\mathrm{hot}$) can still be easily
rotated by an angle $\alpha$ with respect to the pinned magnetization
of the HA$^\mathrm{RT}$ layer by rotating the sample in an external in-plane field of a desired strength.

\begin{figure}[h]
	\includegraphics[width=0.95\columnwidth]{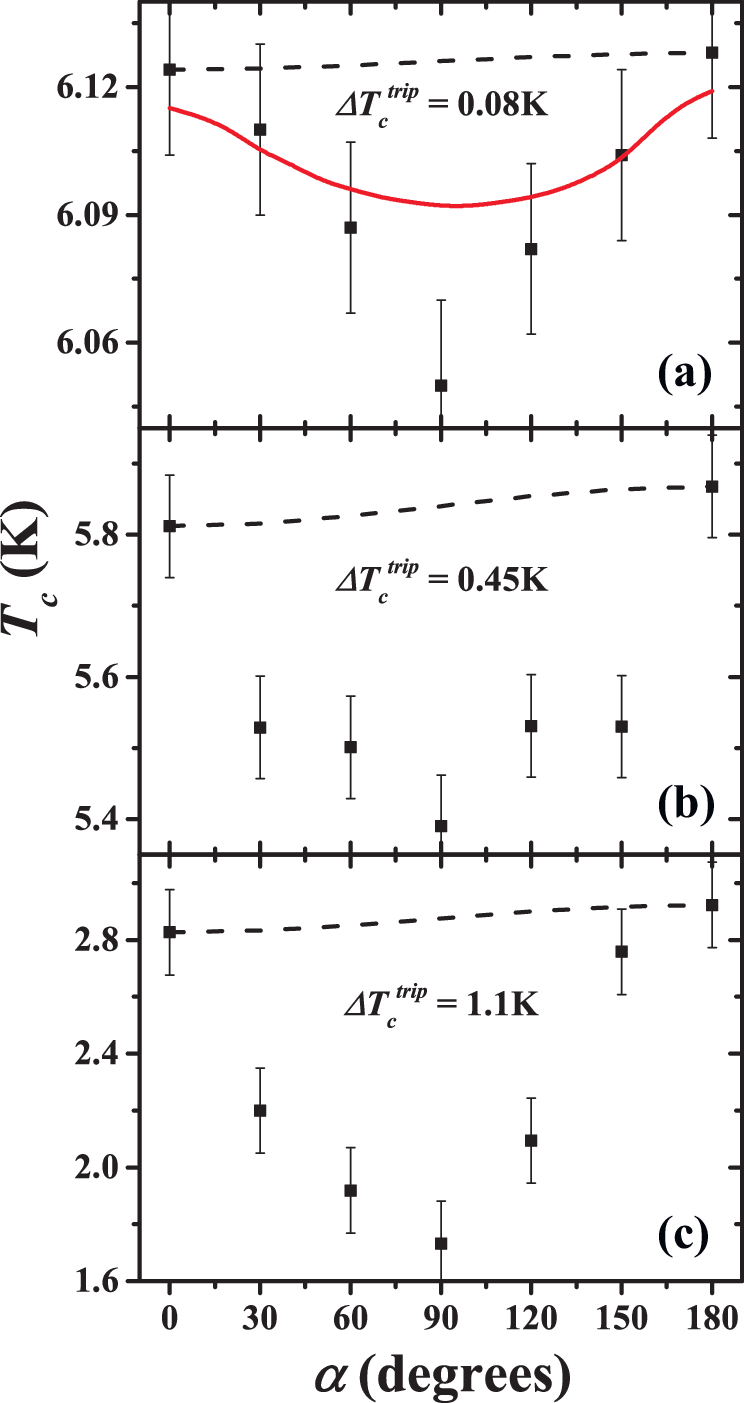}
	\caption{Dependence of $T_c$ on the angle $\alpha$ between the
		direction of the cooling field used to fix the direction of the magnetization of the HA$^\mathrm{RT}$ layer and the applied magnetic
		field that rotates the magnetization of the HA$^\mathrm{hot}$ layer:
        (a)~for the HA$^\mathrm{hot}$(20nm)/\allowbreak Al(4nm)/\allowbreak HA$^\mathrm{RT}$(2nm)/\allowbreak Al(1.2nm)/\allowbreak Pb(60nm) sample
		at the applied magnetic field $H_0 = 1$\,kOe;
        (b)~and (c)~for the HA$^\mathrm{hot}$(20nm)/\allowbreak Al(4nm)/\allowbreak HA$^\mathrm{RT}$(5nm)/\allowbreak Al(1.2nm)/\allowbreak Pb(60nm) sample at the applied magnetic fields
		$H_0 = 2$ and $4$\,kOe, respectively.
		Solid line, theoretical curve with the parameters presented in Sec.~\ref{sec:Discussion};
        dashed lines, reference curves (see Sec.~\ref{sec:Discussion} for details).}
	\label{fig3}
\end{figure}

The optimal performance of the prepared heterostructures as SSVs with regard to the magnitudes of $\Delta T_c^\mathrm{trip}$ and $\Delta T_c^\mathrm{full}$ was achieved for the sample HA$^\mathrm{hot}$(20nm)/\allowbreak Al(4nm)/\allowbreak HA$^\mathrm{RT}$(5nm)/\allowbreak Al(1.2nm)/\allowbreak Pb(60nm).
Other samples from the studied set also demonstrated the SSV effect but with lower values of these parameters.
Figure~\ref{fig3} shows the  dependence of $T_c$ on the angle $\alpha$ between the magnetizations of the F1 and F2 layers measured in different external magnetic fields for two selected samples. All dependences exhibit the characteristic minimum of $T_c$
near the orthogonal PP orientation ($\alpha =90^\circ$) which according to the theory in Ref.\ \cite{Fominov}, indicates the generation of the LRTC of the superconducting condensate.

\begin{figure}[h]
	\includegraphics[width=1\columnwidth]{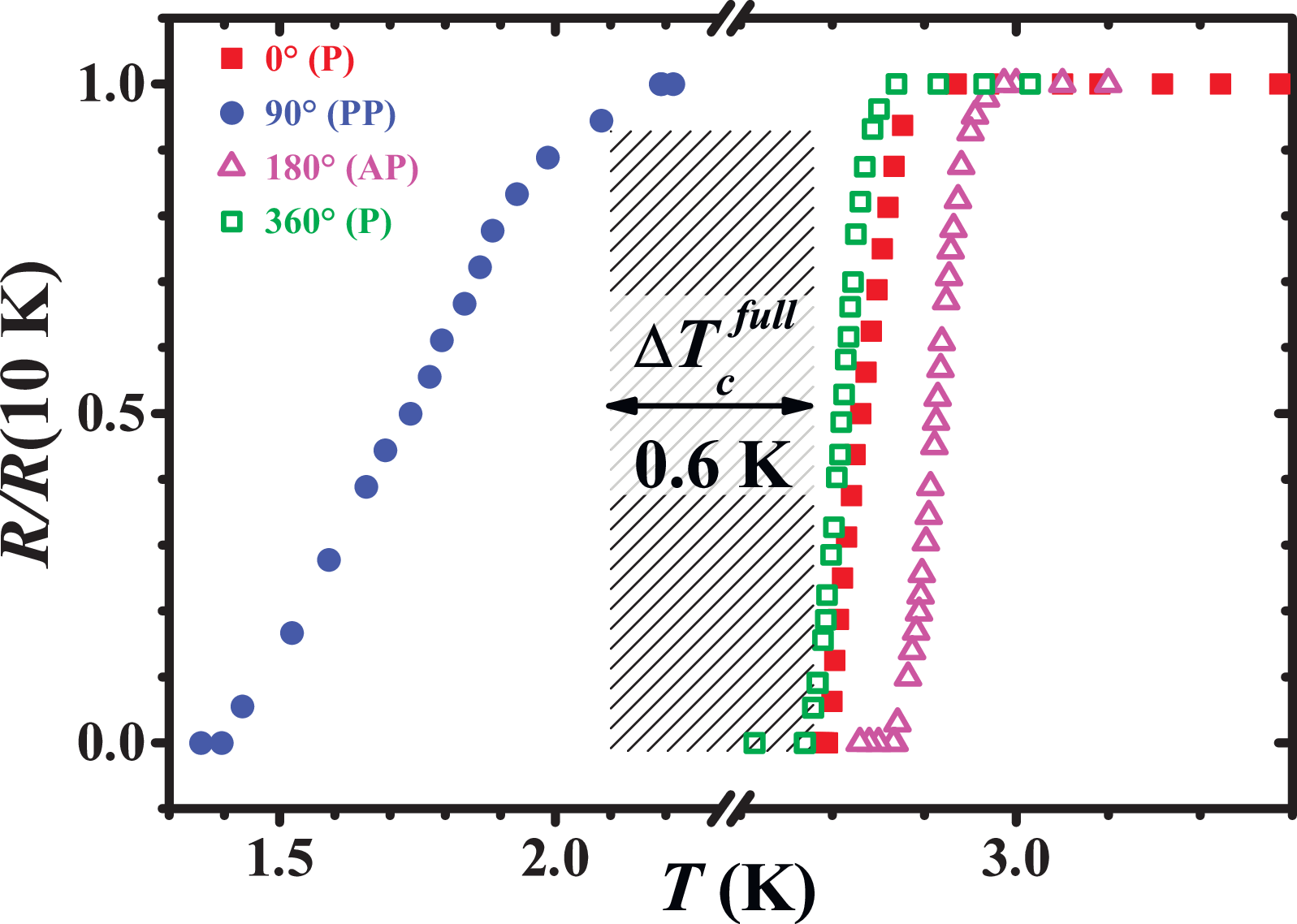}
	\caption{
	Superconducting transition curves for the P ($\alpha=0^\circ, 360^\circ$), AP ($\alpha=180^\circ$) and PP ($\alpha=90^\circ$) 
	configuration of the cooling field used to fix the direction of the magnetization of the HA$^\mathrm{RT}$ layer and the applied magnetic
		field $H_0 = 4$\,kOe that rotates the magnetization of the HA$^\mathrm{hot}$ layer for the
		sample HA$^\mathrm{hot}$(20nm)/\allowbreak Al(4nm)/\allowbreak HA$^\mathrm{RT}$(5nm)/\allowbreak Al(1.2nm)/\allowbreak Pb(60nm).
        The shaded area marks the operational temperature window of the SSV. The three transition curves produce the three points, $T_c(0^\circ)$, $T_c(90^\circ)$ and $T_c(180^\circ)$ in Fig.~\ref{fig3}(c).}
				
	\label{fig4}
\end{figure}

Figure~\ref{fig4} depicts the characteristic superconducting transitions curves for the sample showing the largest spin-triplet SSV effect that  correspond to the data points  P($\alpha=0^\circ, 360^\circ$), AP ($\alpha=180^\circ$) and PP ($\alpha=90^\circ$) in Fig.~\ref{fig3}(c). Here, P, AP and PP are the mutual configurations of the cooling field used to fix the direction of the magnetization of the HA$^\mathrm{RT}$ layer and the applied magnetic
field $H_0 = 4$\,kOe  that rotates the magnetization of the HA$^\mathrm{hot}$ layer. The curves in Fig.~\ref{fig4} demonstrate the magnitude of $\Delta T_c^\mathrm{trip} = T_c(\alpha=0^\circ)-T_c(\alpha=90^\circ)$ of more than 1\,K, the operational temperature window $\Delta T_c^\mathrm{full}$ of about 0.6\,K which is denoted by a shaded area in this Figure, and also the full ordinary SSV effect $\Delta T_c = T_c(\alpha=180^\circ)-T_c(\alpha=0^\circ)$ $\sim$ 85 mK. The superconducting transitions curves for $\alpha=0^\circ$ and $360^\circ$ practically coincide which confirms the full control over the direction of the  magnetization of the F2 layer in the $\alpha = 0^\circ \rightarrow \alpha = 360^\circ$ rotational cycle.


We note that the measurements of the SSV effect presented in Figs.~\ref{fig3} and \ref{fig4} were performed at operational magnetic fields exceeding the coercive field of the individual, single layer of the 
room temperature Heusler alloy (HA$^\mathrm{RT}$, i.e., F2 layer), the magnetization curve of which is shown in Fig.~\ref{fig2}. However, it is plausible that the F2 layer sandwiched in the SSV heterostructure 
between two Al layers gains additional interface anisotropy as compared to the F2 single layer. That under these conditions the F2 layer remains pinned is evident from the observed full periodicity of the 
$T_c$($\alpha$) dependence by rotating the external in-plane field by 360$^\circ$ (Fig.~\ref{fig4}). 
Also, the type of the $T_c$($\alpha$) is independent of the operational field strength, as expected for the robustly
pinned magnetization of the F2 layer. At all measured fields $T_c$($\alpha$) shows a pronounced minimum centered at the orthogonal $\alpha =90^\circ$ orientation that corresponds to the spin-triplet SSV
effect (Fig.~\ref{fig3}). In fact, all studied heterostructures demonstrate also the ordinary SSV effect with $T_c$($\alpha =180^\circ) > T_c$($\alpha =0^\circ$, 360$^\circ$) and a typical for the ordinary SSV effect magnitude $\Delta T_c$ of several tenths mK, which suggests that the magnetization of the F2 layer remains pinned under the used operational conditions (Fig.~\ref{fig4}). It should also be noted that coercivity for a specific direction of the external field does not contain direct information about anisotropy which comes into play when the field is gradually rotated. Strong anisotropy could possibly pin the direction of magnetization in the F2 layer even at large rotating fields. 
Note that gradual rotation of the external field with constant absolute value by 180$^\circ$ can have different effect on our samples than a simple inversion of the field which implies its constant direction and varying the absolute value. In the latter case, the magnetization of the F2 layer is expected to reverse after passing through a demagnetized domain state at the coercive field. In contrast, the rotating field tries to rotate the saturated magnetization without reducing its absolute value. In this case, anisotropy can hinder such rotation and pin the magnetization.
This observation is in agreement with previous experiments \cite{Aarts2015,Kamashev2019,Kamashev20191} in which the $T_c$($\alpha$) dependence was not just observed but even enhanced when the rotating field was increased above the coercive fields of both F layers.

\section{Discussion}
\label{sec:Discussion}

If the LRTC were absent, one would expect a smooth variation of $T_c$ of the SSV structures upon the in-plane rotation of the applied magnetic field and thus of the magnetization of the F1 (HA$^\mathrm{hot}$) layer from the P ($\alpha = 0^\circ$) to the AP ($\alpha = 180^\circ$) orientation. Such an imaginary angular dependence, hereafter called the reference dependence $T_c^\mathrm{(ref)}(\alpha)$, can be represented by a simple interpolation formula $T_c^\mathrm{(ref)}(\alpha)=T_c^\mathrm{P} \cos^2 (\alpha/2) + T_c^\mathrm{AP} \sin^2 (\alpha/2)$ (dashed lines in Fig.~\ref{fig3}) \cite{Leksin2016}.
Thus, a deep minimum of the experimental $T_c(\alpha)$ dependence at the PP ($\alpha = 90^\circ$) configuration (Fig.~\ref{fig3}) implies the significance of spin-triplet superconducting correlations in our samples.
Such behavior is observed at various external magnetic fields such as  $H_0 = 2$ or 4\,kOe.
At the same time, the rather high overall values of $T_c$ in Fig.~\ref{fig3} are probably due to the use of a weak ferromagnet HA$^\mathrm{RT}$ as the F2 layer (which by itself is not too detrimental for superconductivity at such small thickness).

We now discuss our fitting procedure;
the definitions of all the fitting parameters are given in Ref.\ \cite{Kamashev20191}. Similarly to our previous works \cite{Leksin2015,Kamashev20191}, we estimate the values of the coherence lengths $\xi$ from transport measurements in the normal state and the value of $T_{cS} = 7.2$\,K.
The residual resistivity of the Pb layer is $\rho_{\mathrm{S}} = 1.2$\,$\mathrm{\mu}\Omega$\,cm, hence $\xi_\mathrm{S} \approx 46$\,nm.
The residual resistivities of the Heusler layers are $\rho_{\mathrm{F}} = 135$\,$\mathrm{\mu}\Omega$\,cm (they are almost temperature independent up to room temperature). We do not have information about the Fermi parameters of the Heusler layers; however, since its resistivity is two orders of magnitude larger than in the case of Pb, we can expect the corresponding mean free paths $l_\mathrm{F}$ to be two orders of magnitude smaller, and the coherence lengths $\xi_\mathrm{F}$ to be an order of magnitude smaller.
Due to almost identical transport properties of the two F layers, we take $\xi_\mathrm{F1} = \xi_\mathrm{F2}$, hence the materials-matching interface parameter $\gamma_\mathrm{FF} = 1$.
At the same time, the difference of the transport properties of the F2 and S layers leads to the estimate $\gamma_\mathrm{FS} \simeq \xi_{\mathrm{F}2}/\xi_\mathrm{S}$ (under the simplest assumption of equal normal-metallic densities of states and the Fermi velocities in the F2 and S layers).
Taking into account the difference in the DSP of the two F layers, we expect the ratio of the exchange fields $h_1/h_2 \simeq 2-3$.
Transparencies of the interfaces are expected to be good, so we take the interface transparency parameter $\gamma_{b\mathrm{FF}} =0$ (which corresponds to an ideal interface).

The full set of parameters employed in the fitting of Fig.~\ref{fig3}(a) is as follows:
$T_{cS} = 7.2$\,K,
$\xi_\mathrm{S} = 46$\,nm,
$\xi_\mathrm{F1} = \xi_\mathrm{F2} =8$\,nm,
$\gamma_\mathrm{FF} = 1$,
$\gamma_\mathrm{FS} = 0.1$,
$h_1=0.3$\,eV,
$h_2=0.1$\,eV,
$\gamma_{b\mathrm{FF}} =0$,
and $\gamma_{b\mathrm{FS}} =1.13$.

The values of $\xi_\mathrm{F1}$, $\xi_\mathrm{F2}$, $\gamma_\mathrm{FS}$, and the $h_1/h_2$ ratio are consistent with the estimates discussed above.
Their specific values were chosen in order to provide the required nonmonotonic dependence of $T_c(\alpha)$. This behavior is sensitive to the interference effects due to the oscillating nature of the proximity-induced condensate in the ferromagnetic part of the structure, and this guides our choice of the parameters.
Finally, the above value of the interface transparency parameter $\gamma_{b\mathrm{FS}}$ was taken in order to provide the required overall level of $T_c$ values.
Since one can estimate $\gamma_b \sim (l_\mathrm{F}/\xi_\mathrm{F})(1-t_b)/t_b$ in terms of the effective interface transparency $t_b$, this implies that $t_b$ is notably reduced in comparison to unity (that would correspond to the ideally transparent interface). At the same time, even lower values such as $t_b\simeq 0.1$ are not unexpected at a nominally transparent interface between materials with essentially different band structures of the adjacent layers. This difference leads to effective reflectivity of the interface even in the absence of an interface potential barrier.

As evidenced by Fig.~\ref{fig3}(a), our theory is in reasonable agreement with experiment, and indeed captures the main qualitative feature, the minimum of $T_c$ at the PP configuration ($\alpha = 90^\circ$).
At the same time, we cannot expect our theoretical fit to be fully quantitative.
The main reason is that the theory is not suited to the case of strong ferromagnets with significant DSP.
However, although the Usadel equations that we employ
are formally not valid in this case, it
turns out that often they can describe experimental results
surprisingly well (see, e.g., Ref.\ \cite{Moraru2006a}).

We perform theoretical fitting only in Fig.~\ref{fig3}(a) because the results in Figs.~\ref{fig3}(b) and (c) demonstrate additional physical effects not taken into account in our theory.
In those cases, the effect of the external magnetic field cannot be simply reduced to the rotation of the magnetization.

Our experimental results indicate that the value of $\Delta T_c^\mathrm{trip}$ increases with increasing the strength of the external magnetic field, e.g., from $\Delta T_c^\mathrm{trip} \approx 0.45$\,K at $H_0 = 2$\,kOe in Fig.~\ref{fig3}(b) to $\Delta T_c^\mathrm{trip} \approx 1.1$\,K at $H_0 = 4$\,kOe in Fig.~\ref{fig3}(c).
Notably, at the same time the overall values of $T_c$ are getting suppressed, as can be clearly seen by comparing panels (a), (b), and (c) of Fig.~\ref{fig3} in the case of the P configuration ($\alpha=0^\circ$). This can be interpreted as being due to the suppression of the critical temperature of the thin S film in the parallel in-plane magnetic field. It is therefore somewhat surprising that despite the overall suppression of $T_c$, the amplitude of the variation of $T_c$ as a function of the angle between the F1 and F2 magnetizations increases.

The reason for this striking effect is yet to be elucidated.
Presumably this could be due to the not fully saturated magnetizations in the F part of the structure either because of the persistence of domains or surface magnetic inhomogeneities. An additional magnetizing of an F layer by the increasing external magnetic field could then
make the SSV effect more pronounced.
Another side of the same mechanism could be effective spin-flip scattering due to magnetic inhomogeneities \cite{Ivanov2009} that leads to damping of superconducting correlations and is thus detrimental to interference effects caused by the oscillating nature of the proximity effect in ferromagnets. Better ordering of magnetic moments would suppress the spin-flip scattering and thus enhance the SSV effect.

We note that similar increase of the SSV effect by an external magnetic field has been earlier observed in Refs.\ \cite{Aarts2015,Kamashev2019,Kamashev20191}.
We also note that the external field applied in order to rotate magnetization of the F1 layer in the cases of Figs.~\ref{fig3}(b) and (c) is clearly larger than the coercive field of the F2 layer as well.
That under these conditions the F2 layer remains pinned is evident from the observed full periodicity of the $T_c(\alpha)$ dependence by rotating the external in-plane field by 360$^\circ$. It is plausible that the F2 layer sandwiched in the SSV heterostructure between two Al layers gains additional interface anisotropy as compared to the F2 single layer the magnetization loop of which is shown in Fig.~\ref{fig2}. Also, coercivity for a specific direction of external field does not contain direct information about anisotropy which comes into play when the field is gradually rotated. Strong anisotropy could possibly pin the direction of magnetization in the F2 layer even at large rotating fields.
This observation is in agreement with previous experiments \cite{Aarts2015,Kamashev2019,Kamashev20191} in which the $T_c(\alpha)$ dependence was not just observed but even enhanced when the rotating field was increased above the coercive fields of both F layers.

\section{Conclusions}

In summary, we have exploited the tunability of the magnetic properties of the Heusler alloy Co$_2$Cr$_{1-x}$Fe$_x$Al$_{y}$ (HA) for its use in a dual role in the superconducting spin valve heterostructure F1/F2/S (S = Pb). While fabricating this structure, we first deposited the HA film at a high temperature of $\sim 700^\circ$ to play the role of the practically half-metallic F1 layer and, in the next step, repeated the deposition of HA at room temperature to prepare the weakly ferromagnetic F2 layer. The so prepared heterostructures HA$^\mathrm{hot}$/Al/HA$^\mathrm{RT}$/Al/Pb with Al interlayers feature a pronounced long-range triplet component of the superconducting condensate at the orthogonal orientation of the magnetizations of the HA$^\mathrm{hot}$ and HA$^\mathrm{RT}$ layers manifesting in a very strong suppression of the superconducting transition temperature $\Delta T_c^\mathrm{trip}$ of more than 1\,K for this orientation. Furthermore, increasing the strength of the in-plane magnetic field that rotates the magnetization of the F1 layer up to 4\,kOe  enabled us to expand the operational temperature window of the SSV to
$\Delta T_c^\mathrm{full} \sim 0.6$\,K.
These results correlate to some extent with the results obtained earlier in Refs.\ \cite{Aarts2015,Kamashev2019,Kamashev20191,Kamashev2020}. However, in those previous works the ferromagnetic part of the SSV comprised a half-metallic ferromagnetic and a more conventional ferromagnetic layer. In the present paper we realized a conceptually new design of an SSV where we fully exploited the flexibility of the Heusler alloy Co$_2$Cr$_{1-x}$Fe$_x$Al$_y$ by uniting for the first time its hot version HA$^\mathrm{hot}$ as an F1 layer and its room temperature version HA$^\mathrm{RT}$ as an F2 layer in the very same F1/F2/S heterostructure which enabled us to achieve a much larger magnitude of the effects, especially of the operational temperature window.

The achieved values of $\Delta T_c^\mathrm{trip}$ and $\Delta T_c^\mathrm{full}$ set  benchmarks for the design of the superconducting spin valves as the prototype elements for applications in  superconducting spintronics. They demonstrate the by far not yet exhausted potential for the optimization of ferromagnetic materials in an F1/F2/S SSV structure together with the simultaneous simplification of its fabrication.
The observed boosting of the SSV effect by increasing the strength of the rotating external magnetic field on the background of the overall suppression of $T_c$ calls for further experimental and theoretical studies for a better understanding of the underlying physics of this phenomenon.

\acknowledgments

The work of A.A.K.\ and N.N.G.\ concerning the preparation of the samples was funded by the Russian Scientific Foundation according to research project No.\ 21-72-20153.
The work of A.A.K., N.N.G., A.A.V., and I.A.G.\ concerning the investigation of the triplet spin-valve effect was financially supported by the government assignment for
FRC Kazan Scientific Center of Russian Academy of Sciences.
Ya.V.F.\ was supported by the Foundation for the Advancement of Theoretical Physics and Mathematics ``BASIS''.

\end{document}